\documentclass[pra,twocolumn,showpacs]{revtex4}

\usepackage{graphicx}
\usepackage{amsmath}
\usepackage{times}
\usepackage{dcolumn}

\begin{document}

\title[Phonons in freely expanding BEC]{The fate
of phonons in freely expanding Bose-Einstein condensates}

\author{C. Tozzo}
\affiliation{Dipartimento di Fisica, Universit\`a di Trento, and 
Istituto Nazionale \\ per la Fisica della Materia, BEC-INFM Trento,
I-38050 Povo, Italy }

\author{F. Dalfovo}
\affiliation{Dipartimento di Matematica e Fisica, Universit\`a
Cattolica, Via Musei 41, 25121 Brescia and \\
Istituto Nazionale per la Fisica della Materia, Unit\`a di Brescia
and BEC-INFM Trento}

\date{\today}

\begin{abstract}

Phonon-like excitations can be imprinted into a trapped Bose-Einstein 
condensate of cold atoms using light scattering. If the condensate 
is suddenly let to freely expand, the initial phonons lose their 
collective character by transferring their energy and momentum to 
the motion of individual atoms. The basic mechanisms of this 
evaporation process are investigated by using the Gross-Pitaevskii 
theory and dynamically rescaled Bogoliubov equations. Different 
regimes of evaporation are shown to occur depending on the phonon 
wavelength. Distinctive signatures of the evaporated phonons are 
visible in the density distribution of the expanded gas, thus 
providing a new type of spectroscopy of Bogoliubov excitations.   

\end{abstract}


\maketitle

\section{Introduction}
\label{sec:intro}

Bose-Einstein condensates of cold atoms represent a remarkable testing 
ground for theories of weakly interacting bosons. The fact that the 
actual condensates are spatially confined and inhomogeneous makes these 
systems even more interesting. From one hand, the Bogoliubov theory 
of uniform gases \cite{bogoliubov} can still be applied in certain 
limits, when the system can be viewed as ``locally uniform". Such a 
local density approach has been used, for instance, to characterize 
the response of a condensate to light scattering processes 
\cite{stamperkurn,vogels,steinhauer1,ozeri,katz,zambelli,brunello}.
In the same spirit, properties of the uniform gas can also be extracted 
from a trapped condensate by reducing the effects of inhomogeneous 
broadening as suggested in Ref.~\cite{gershnabel}. On the other hand,
finite size effects are interesting by themselves, giving rise to a
new type of Bogoliubov excitations with spatially varying 
quasiparticle amplitudes, such as the low frequency collective 
oscillations of the whole condensate (see \cite{rmp} and references
therein) and the axial phonons of elongated condensates, exhibiting 
a multibranch spectrum \cite{steinhauer2,tozzo}. 

In most of the experiments with excited condensates,
the observations are performed after switching-off the confining 
potential and letting the condensate to expand.  The comparison 
between theory and experimental data often ignores the dynamics of 
the expansion. This is justified if one looks for quantities that 
are conserved during the expansion like, for instance,  
the total momentum of the condensate. 
However, the expansion of a condensate initially dressed with 
phonon-like modes is interesting from both the conceptual and 
experimental viewpoints. Conceptually, the behavior of an excited 
state that starts as a quasiparticle and evolves into a particle
is a remarkable and nontrivial example of quantum process, which 
shows some similarities with the quantum evaporation process at the
surface of superfluid helium \cite{helium} and with the evolution
of two-level systems with nonhermitian Hamiltonian \cite{twolevel}. 
From the experimental viewpoint, on the other hand, the characterization
of the observable density and velocity distributions of the 
expanded gas in terms of specific initial configurations is 
important in order to use those distributions as a probe of 
in-trap quasiparticles.  

In this paper we use the Gross-Pitaevskii (GP) theory to investigate 
the expansion of elongated axially symmetric condensates. We assume
the initial configuration to be a stationary trapped condensate
at zero temperature. The condensate is excited by populating some 
quasiparticle states and, then, let to expand. 

In section II, we first present the results of numerical simulations 
based on the direct integration of the time dependent Gross-Pitaevskii 
equation in the case of an elongated condensate similar to the one of 
recent experiments \cite{steinhauer1,ozeri,steinhauer2}. We show 
that the expansion exhibits quite different behaviors depending on 
the wavelength of the initial phonons: i) long wavelength 
phonons remain inside the expanding condensate in the form of 
density modulations; ii) short wavelength phonons are converted 
into a separate cloud of excited atoms moving out of the condensate. 
In both cases, the final density distribution shows nontrivial 
features, such as a radial distortion of the density modulations
in the condensate and a ``shell" structure of the released-phonon
cloud.  

In section III, we use the simplified geometry of an infinite 
cylindrical condensate in order to get a deeper insight, pointing 
out the different role of radial and axial degrees of freedom. In 
this case, we numerically solve the Bogoliubov equations for the 
amplitudes of in-trap quasiparticles. Then we follow
the expansion of a condensate initially dressed with one of these 
quasiparticles by using the scaling properties of the GP equation in 
two-dimensions and solving rescaled Bogoliubov-like equations. This 
approach allows us to characterize the behavior of the excitations 
during the expansion, at different levels of approximation, and to 
find the relevant timescales for the evaporation process. Interesting
results can also be obtained by averaging out the slow radial motion 
of the excitations in the rescaled coordinates. The problem is thus 
mapped into the evolution of a quasiparticle in a uniform gas with a
decreasing time-dependent density. In section IV, we show that this
process can be either adiabatic (conversion of a quasiparticle into a 
single particle with the same momentum) or non-adiabatic depending on 
the phonon wavelength and on the chemical potential of the gas. 

In section V, we compare the predictions of the rescaled Bogoliubov
equations for the infinite cylinder with the results of GP simulations 
for elongated condensates. The comparison is instructive and allows one 
to relate the properties of in-trap quasiparticles with observable 
features of the expanded gas. In particular, we discuss the axial 
motion of density modulations, the radial scaling of nodal lines, 
the conditions for the appearance of a separate released-phonon cloud
and for the adiabaticity of the quasiparticle evaporation process.  

The expansion of condensates with long wavelength phonons was 
previously studied in Ref.~\cite{dettmer}, where random phase 
fluctuations were included in the configuration of a very elongated 
condensate at the beginning of the expansion, in order to simulate 
the behavior of a quasi-condensate with thermal excitations. Low energy 
phonons that expand in the hydrodynamic regime were also considered 
in \cite{fedichev1} as due to ``quantum vacuum" phase fluctuations.
Instead of investigating the effects of thermal and/or quantum 
fluctuations, we are here considering situations where the initial 
excitations are imprinted in a controllable way and characterized 
by looking at the shape of the expanded gas, thus allowing for a 
new type of spectroscopic studies of Bogoliubov quasiparticles.

\section{GP numerical simulations for elongated condensates}
\label{section1}

We simulate a process of quasiparticle creation and expansion
in a realistic axially symmetric, cigar-shaped, condensate. 
The starting point is the time dependent GP equation for the order 
parameter $\Psi (x,y,z,t)$ of a condensate of $N$ bosonic atoms
of mass $m$ \cite{pitaevskii1,rmp}: 
\begin{equation}
i \hbar \partial_t \Psi = 
\left( - \frac{ \hbar^2 \nabla^2 }{ 2m } + V +gN |\Psi|^2 
\right) \Psi \; ,
\label{eq:TDGP}
\end{equation}
where $g=4 \pi \hbar^2 a/m$, and $a$ is the $s$-wave scattering 
length that we assume to be positive. The order parameter is here
normalized according to $\int d{\bf r} |\Psi|^2 = 1$.

Let us first take the harmonic trapping potential in the form
\begin{equation}
V (\rho,z) = V_{\rm trap} (\rho,z) 
= (1/2) m \omega_\rho^2 (\rho^2 + \lambda^2 z^2)
\label{eq:trap}
\end{equation}
where $\rho = [x^2 + y^2]^{1/2}$ and $\lambda=\omega_z/\omega_\rho$. 
The ground state can be found as the stationary solution of 
Eq.~(\ref{eq:TDGP}). In practice we map the order parameter on a 
$N_\rho \times N_z$ grid (typically, $64 \times 1024$ points) and 
propagate it in imaginary time with an explicit first order algorithm 
starting from a trial configuration, as described in Ref.~\cite{ds}. 
If $\lambda < 1$ the condensate at equilibrium is a prolate ellipsoid.  

\begin{figure}[htb]
\includegraphics[width=3.3in]{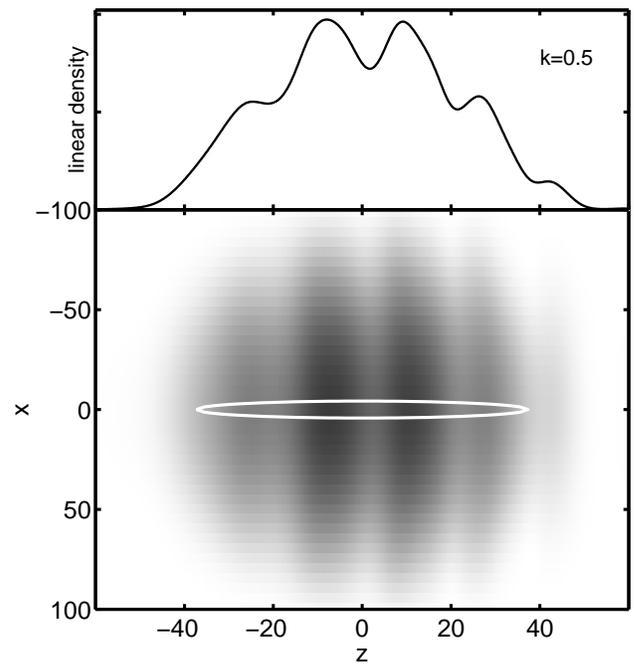}
\caption{ \label{fig:GPdensity1}
Column (below) and linear (above) density, in arbitrary units,
of a condensate subject to a Bragg pulse with $k=0.5 a_\rho^{-1}$, 
$\omega= 1.1 \omega_\rho$, $V_B= 0.2 \hbar \omega_\rho$ and 
$t_B=4 \omega_\rho^{-1}$, and a subsequent free expansion of 
$t=25 \omega_\rho^{-1}$. The white ellipse is the shape of the 
condensate at $t=0$, at the beginning of the expansion. Distances
are in units of $a_\rho= [\hbar/(m\omega_\rho)]^{1/2}$.}
\end{figure}

\begin{figure}[htb]
\includegraphics[width=3.3in]{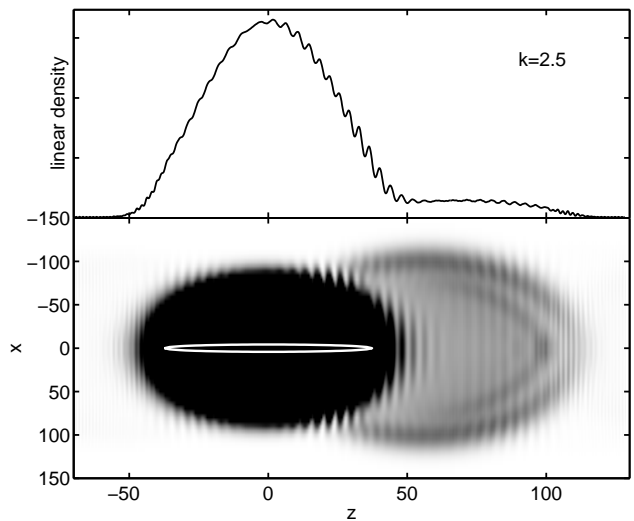}
\caption{ \label{fig:GPdensity2}
Same as before, but for a Bragg pulse with $k=2.5 a_\rho^{-1}$,
$\omega= 4.7 \omega_\rho$, $V_B= 1 \hbar \omega_\rho$ and 
$t_B=1 \omega_\rho^{-1}$.}
\end{figure}

Excitations can be created in the condensate by using light
(Bragg) scattering \cite{stamperkurn,vogels,steinhauer1,ozeri,katz}. 
Within the GP theory, this is accounted for by adding an extra 
term to the external potential, that becomes \cite{blakie}
\begin{equation}
V(\rho,z,t)= V_{\rm trap}(\rho,z) + V_B \cos(k z-\omega t) 
\label{eq:vbragg}
\end{equation}
for $- t_B < t <0$ and assuming the momentum transfer to be along 
$z$. We numerically integrate the GP equation~(\ref{eq:TDGP}) 
by propagating the order parameter in real time, starting from the 
ground state configuration at $t=-t_B$. We split the propagation 
into the axial and radial parts.  The former is obtained by using 
a fast Fourier transform algorithm to treat the kinetic energy 
term, while the latter is  performed with a Crank-Nicholson 
differencing method, as in Refs.~\cite{brunello,steinhauer2,tozzo}. 
 
Now, let's suppose that both the Bragg pulse and the external 
trapping potential  are switched-off at $t=0$. 
The condensate freely expands and the 
expansion can still be studied by solving the GP equation. We use
the same algorithm as before with the only difference that we 
rescale all distances during the expansion in order to keep the 
box size and the number of grid points the same as for the
trapped condensate.  For that we use a dynamically rescaled GP 
equation already introduced in \cite{modugno1,modugno2}.

For our simulations we choose the parameters of the condensate 
used in the experiments of Nir Davidson and co-workers 
\cite{steinhauer1,ozeri,katz,steinhauer2}, made of $N=10^5$ atoms 
of $^{87}$Rb in a trap with $\omega_\rho= 2\pi (220$Hz$)$ and 
$\lambda=0.114$. We plot distances in units of the harmonic 
oscillator length $a_\rho= [\hbar/(m\omega_\rho)]^{1/2}$, time in 
units of $\omega_\rho^{-1}$ and energies in units of $\hbar 
\omega_\rho$. The chemical potential is $\mu = 9.1 \hbar 
\omega_\rho$. Several condensates in current experiments have 
similar parameters. 

Typical results are shown in Figs.~\ref{fig:GPdensity1} and
\ref{fig:GPdensity2}, where we plot the column density, $\int\! 
dy |\Psi(x,y,z)|^2$, and the linear density, $\int\! dx dy 
|\Psi(x,y,z)|^2$, for two sets of parameters of the 
Bragg pulse. The values are chosen such that the lowest 
longitudinal modes are resonantly excited. Moreover, both 
values of $k$ are smaller than $\xi^{-1} \simeq 3 a_\rho^{-1}$, 
where $\xi$ is the healing length. This implies that in both cases 
the excitations are collective phonon-like Bogoliubov 
quasiparticles.  

There is a dramatic difference between the density distributions 
for the two values of $k$. At low $k$ all atoms remain within the 
expanding condensate that exhibits an evident density modulation, 
while at high $k$ an extra cloud of excited atoms separates out 
in the positive $z$ direction.   

\begin{figure}[htb]
\includegraphics[width=3.3in]{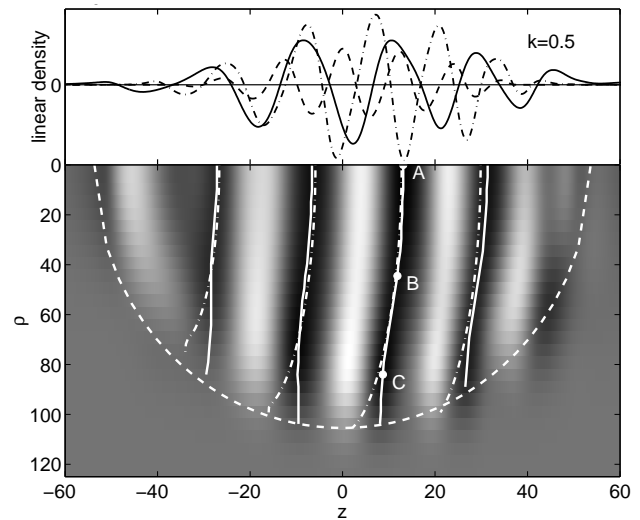}
\caption{ \label{fig:GPdeltan1}
Lower part: the quantity $\Delta n(\rho,z)$ at $t=25 \omega_\rho^{-1}$ 
is plotted, in arbitrary units, for the same simulation of 
Fig.~\protect\ref{fig:GPdensity1}. Black (white) means positive 
(negative) $\Delta n(\rho,z)$. The dashed line is the position of 
the surface of the expanding condensate. The uniform grey colour
far outside the condensate corresponds to $\Delta n=0$.  Solid lines 
are the position of maxima of $\Delta n(\rho,z)$ (wavefronts). 
Dot-dashed lines are the same wavefronts estimated by means of 
Eq.~(\protect\ref{eq:simplemodel}). Upper part: integrated density
variation $2\pi \int d\rho \rho \Delta n$, calculated at $t=0$ 
(dashed), $10$ (dot-dashed) and $25 \omega_\rho^{-1}$ (solid). } 
\end{figure}

\begin{figure}[htb]
\includegraphics[width=3.3in]{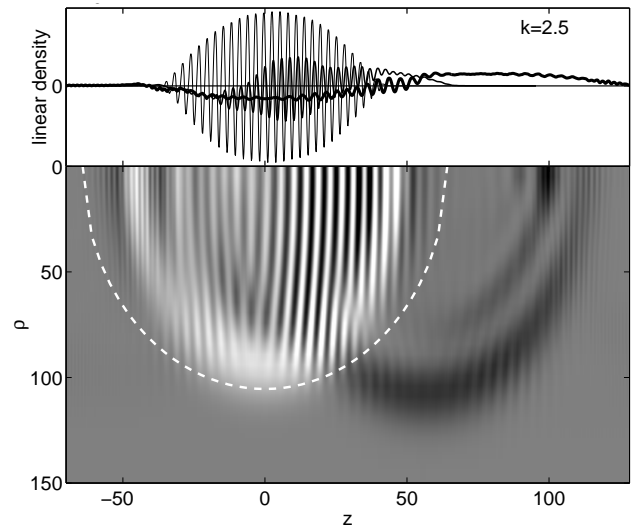}
\caption{ \label{fig:GPdeltan2}
Lower part: the quantity $\Delta n(\rho,z)$ at $t=25 \omega_\rho^{-1}$ 
is plotted, in arbitrary units, for the same simulation of 
Fig.~\protect\ref{fig:GPdensity2}. The meaning of the greyscale is 
the same as in the lower part of Fig.~\protect\ref{fig:GPdeltan1}. 
Upper part: integrated density variation $2\pi \int\! d\rho \ 
\rho \Delta n$, calculated at $t=0$, $10$ and $25 \omega_\rho^{-1}$ 
from thin to thick solid line, respectively. }
\end{figure}

In order to get a better view of the behavior of expanding phonons 
one can repeat each simulation twice for the same condensate with 
and without excitations (Bragg pulse ``on" and ``off") and calculate 
the density difference $\Delta n(\rho,z) = N [ |\Psi_{\rm on}
(x,y,z)|^2 - |\Psi_{\rm off}(x,y,z)|^2]$. When the number of 
quasiparticles is much smaller than $N$, this quantity converges 
to the density variation, $\delta n(\rho,z)$, associated with the 
excitations in the linear response regime.  In Figs.~\ref{fig:GPdeltan1} 
and \ref{fig:GPdeltan2} we show the results for the two  sets of 
parameters of Figs.~\ref{fig:GPdensity1} and \ref{fig:GPdensity2}, 
respectively.   

\begin{figure}[htb]
\includegraphics[width=3.3in]{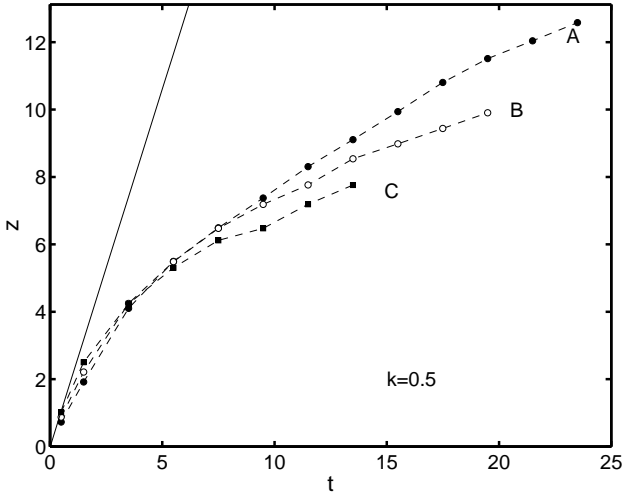}
\caption{ \label{fig:GPfrontspeed}
Position $z(t)$ of a wavefront for $k=0.5 a_\rho^{-1}$ and 
$\rho = 0, 0.4$ and $0.8 R$ (points A, B and C in Fig.~3), 
where $R$ is the radial size of the expanding condensate. 
The thin solid line is $z(t)=ct$, where $c=[g n(0)/(2m)]^{1/2}$ is
the sound velocity in a cylindrical condensate with the same 
central density $n(0)$. }
\end{figure}

At $k=0.5 a_\rho^{-1}$ we observe well defined wavefronts. They 
reflect the density modulations associated with the initial phonon 
at $t=0$ (dashed line in the upper plot). These modulations move 
along $z$ during the expansion. Their velocity depends on both 
$z$ and $\rho$ and is faster for $\rho=0$, where the density is 
higher. This causes each front to be slightly bent. Effects of such 
a bending are also visible in Fig.~\ref{fig:GPdensity1} in the 
form of a small asymmetry of the modulations of both the column and 
linear densities.  In Fig.~\ref{fig:GPfrontspeed} we show the 
position $z(t)$ of a wavefront for $\rho = 0, 0.4$ and $0.8 R$ 
(points A, B and C in Fig.~\ref{fig:GPdeltan1}), where $R$ is radial 
size of the expanding condensate. The initial velocity is very 
close to the sound velocity in a cylindrical condensate, $c= 
[g n(0)/(2m)]^{1/2}$ \cite{zaremba}, having the same density 
profile $n(\rho)$ at $z=0$.  

At $k=2.5 a_\rho^{-1}$ the main feature is the shape of the cloud of 
excited atoms that are moving out of the condensate. By using the 
order parameter $\Psi(x,y,z)$, one can calculate the local contribution
to the total energy and momentum of the system. Most of the energy of
the condensate is, of course, in the kinetic energy of the fast radial 
motion. What is more important, however, is that the initial excitation
energy, i.e., the energy of the excited phonons, is almost completely 
transferred to the lateral cloud, which is moving along $z$ at a velocity 
of the order of $\hbar k/m$.  The shape of this cloud exhibits an 
interesting ``shell" structure with some lobes divided by almost empty 
regions. The appearance of this extra cloud is accompanied by a strong 
suppression of density modulations in the condensate. 

By performing several simulations at different $k$ we find a transition
from the low-$k$ scenario (density modulations within the expanding 
condensate) and the higher $k$ scenario (external cloud of atoms)
occurring in between $k=0.5$ and $1 a_\rho^{-1}$. The experiments so 
far performed with phonons excited by Bragg scattering
\cite{stamperkurn,vogels,steinhauer1,ozeri,katz,steinhauer2}
belongs to the second scenario. Using a tomographic imaging method 
the authors of Ref.~\cite{ozeri} found that the observed 
released-phonon cloud has indeed a nontrivial shell-like shape.
First evidences of the first scenario have also been found in 
recent experiments of the same group \cite{davidson}, where 
density modulations have been observed at low $k$ within the
condensate. Before coming back to the interpretation of our GP
simulations, let us explore the instructive case of an infinite 
cylindrical condensate. 

\begin{figure}[htb]
\includegraphics[width=3.3in]{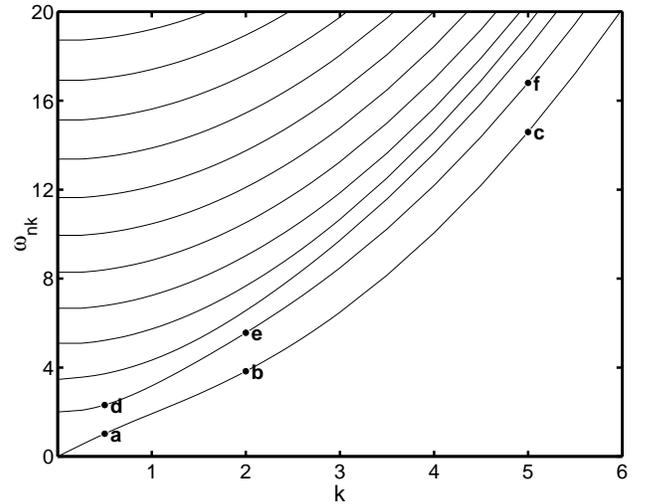}
\caption{ \label{fig:spectrum}
Spectrum of axially symmetric Bogoliubov excitations of a cylindrical 
condensate with $\eta=9.1$. The frequency $\omega_{nk}$, in units of 
the radial trapping frequency $\omega_\rho$, is plotted as a function 
of the axial wavevector $k$, in units of $a_{\rho}^{-1}$. The 
number of radial nodes is $n=0,1,2,\dots$, starting from the lowest 
branch. The quasiparticle amplitudes $u$ and $v$ of the modes
$a, b, c, d, e$ and $f$ are shown in the first column of
Figs.~\protect\ref{fig:uandv1} and \protect\ref{fig:uandv2}.      }
\end{figure}

\section{Bogoliubov excitations in expanding cylindrical condensates}
\label{section2}

\subsection{Bogoliubov excitations in trap}

Here we consider a condensate that is unbound along $z$ and radially 
confined in the harmonic potential $V(\rho)=(1/2) m \omega_\rho^2 
\rho^2$. In this geometry one can better distinguish the different 
roles played by radial and axial degrees of freedom. 
The order parameter of the ground state only depends on $\rho$ and
obeys the stationary GP equation
\begin{equation}
\left[-\frac{\hbar^2 \nabla^2_\rho}{2m} +V(\rho)+ gn_0(\rho)
\right] \Psi_0(\rho)=\mu\Psi_0(\rho)
\label{eq:Psi0}
\end{equation}
where $\mu$ is the chemical potential and $\Psi_0$ is chosen to be 
real and subject to the normalization condition 
$2\pi \int_0^{\infty}\! d\rho \,  \rho  \Psi_0^2 =1$, so that
$n_0(\rho)= (N/L)\Psi_0^2(\rho)$ is the ground state density.
The excited states are plane waves along $z$, with wavevector $k$.
Different branches of such longitudinal waves exist, characterized
by the number of nodes in the radial direction, $n$.  When 
the condensate is weakly excited in one of these modes, its order 
parameter can be written as
\begin{equation}
\Psi (\rho,z,t) =  e^{-i\mu t/\hbar} [ \Psi_0 (\rho) + 
\delta \Psi (\rho,z,t) ]
\end{equation}
where 
\begin{eqnarray}
\delta \Psi (\rho,z,t) = L^{-1/2} &[&  u_{nk}(\rho) 
                 e^{ i (kz - \omega_{nk} t)} \nonumber \\
        &+&  v_{nk}^*(\rho)  e^{-i (kz - \omega_{nk} t)} ] 
\; .
\label{eq:linearized}
\end{eqnarray}
Inserting this expression into Eq.~(\ref{eq:TDGP})
one gets the Bogoliubov equation
\begin{equation}
\left( \begin{array}{cc}
H_\rho   + \frac{\hbar^2 k^2}{2m} & g n_0 (\rho)  \\
-g n_0 (\rho)  & - H_\rho  - \frac{\hbar^2 k^2}{2m}
\end{array} \right)
\left( \begin{array}{c} u_{nk} \\ v_{nk}
\end{array} \right) =
\hbar \omega_{nk} \left( \begin{array}{c} u_{nk} \\ v_{nk}
\end{array} \right)
\label{eq:bog}
\end{equation}
where
\begin{equation}
H_\rho = -\frac{\hbar^2\nabla_\rho^2}{2m} + V(\rho) +
2 g n_0(\rho)- \mu \; .
\label{eq:Hrho}
\end{equation}   
The quasiparticle amplitudes $u_{nk}$ and $v_{nk}$ obey the 
following orthogonality and symmetry relations
\begin{eqnarray}
& \int\! & d\rho \ 2 \pi \rho ( u_{nk} u_{n'k'}^* - 
v_{nk} v_{n'k'}^* )  =  \delta_{nn'} \delta(k-k') \\
& \int\! & d\rho \ \rho (u_{nk} v_{n'k'} - v_{nk} u_{n'k'} ) = 0 
\; .
\label{eq:ortho}
\end{eqnarray}
Equations (\ref{eq:Psi0}) and (\ref{eq:bog}) can be solved 
numerically to get $\Psi_0$, $u_{nk}$, $v_{nk}$ and $\omega_{nk}$ 
\cite{tozzo,fedichev2,muntsa,komineas}. 

\begin{widetext}

\begin{figure}[htb]
\includegraphics[width=5in]{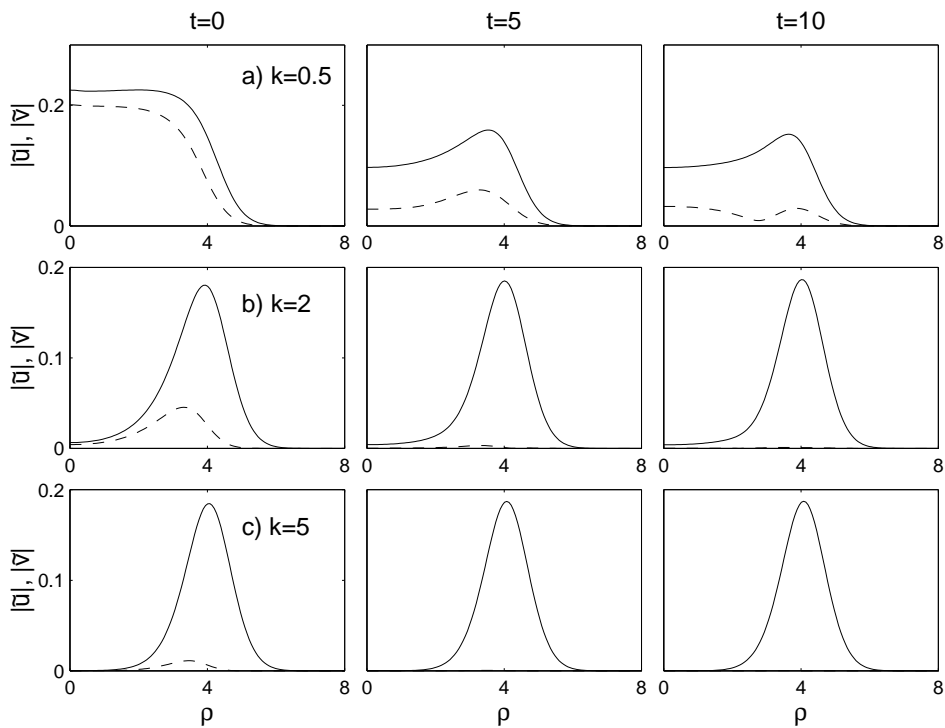}
\caption{ \label{fig:uandv1}
Time evolution of the functions $|\tilde{u}(\rho)|$ (solid lines) 
and $|\tilde{v} (\rho)|$ (dashed lines) obtained from 
Eq.~(\protect\ref{eq:scaledBG1}). At $t=0$ these functions coincide 
with the Bogoliubov amplitudes $u_{0k}$ and $v_{0k}$, solutions of 
Eq.~(\protect\ref{eq:bog}) for the modes (a), (b), and (c) in the 
spectrum of Fig.~\protect\ref{fig:spectrum}. The radial coordinate 
$\rho$ is given in units of $a_\rho$. The Thomas-Fermi radius of the 
condensate is $R= 4.27 a_\rho$.  }
\end{figure}

\end{widetext}

\begin{widetext}

\begin{figure}[htb]
\includegraphics[width=5in]{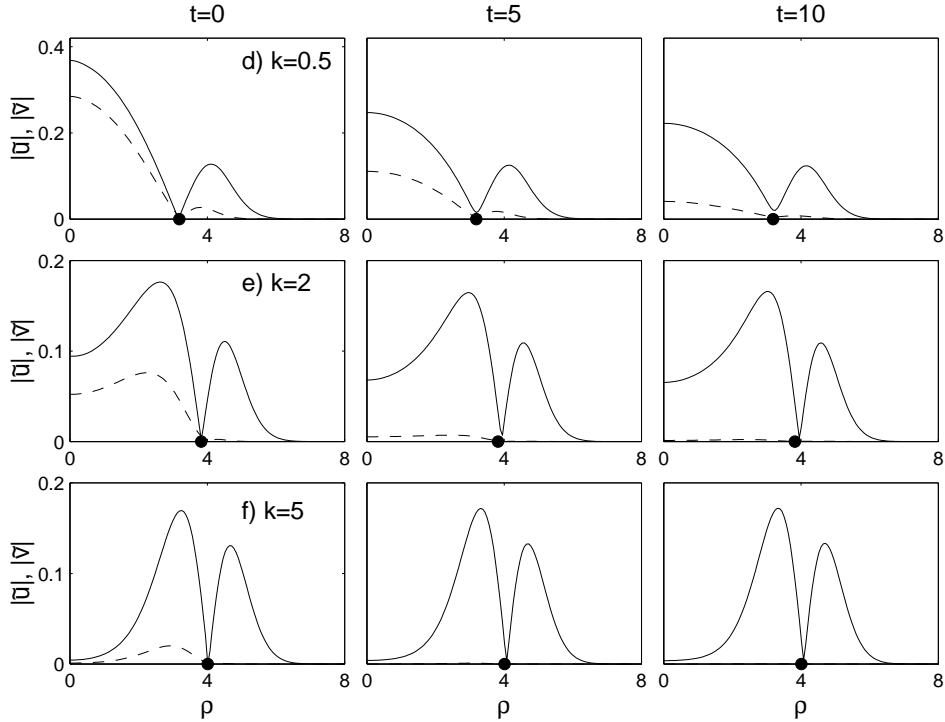}
\caption{ \label{fig:uandv2}
Same as in Fig.~\protect\ref{fig:uandv1}, but for the modes 
(d), (e) and (f). At $t=0$ the functions $|\tilde{u}|$ and $|\tilde{v}|$ 
coincide with the Bogoliubov amplitudes $u_{1k}$ and $v_{1k}$, solutions 
of Eq.~(\protect\ref{eq:bog}) with one radial node. The position of the 
$t=0$ node is shown as a black dot on the horizontal axis.  }
\end{figure}

\end{widetext}

In Fig.~\ref{fig:spectrum} we show the spectrum of axial phonons in
a cylindrical condensate which simulates the behavior of the 
elongated condensate discussed in section II. The relevant parameter 
that characterizes the solutions of the above equations is $\eta \equiv  
\mu_{\rm TF} / (\hbar \omega_\rho) = [ 4 a N/L]^{1/2}$, where $\mu_{\rm 
TF}$ is the chemical potential in the Thomas-Fermi limit. The condensate
in Fig.~\ref{fig:spectrum} has $\eta=9.1$. Similar spectra were  
already discussed in details in Ref.~\cite{tozzo}. The form of the
quasiparticle amplitudes $u_{nk}(\rho)$ and $v_{nk}(\rho)$ for some 
states on the first two branches is shown in Figs.~\ref{fig:uandv1} 
and \ref{fig:uandv2} ($t=0$ plots, first column). The modes on the 
lowest branch have no nodes in the radial direction; those in the first 
branch have one node, and so on. One also notes that the excited 
states have a nonvanishing amplitude in the center of the condensate 
only at low $k$, while at higher $k$ they are mainly located in
the low density region near the surface \cite{rmp,muntsa2}. Finally,
we observe that, as expected, the amplitude $v$ is of the same order 
of $u$ in the phononic regime at small $k$, while it is  negligible 
when $k$ is larger than $\xi^{-1}$. In our case $\xi^{-1} = \eta^{1/2}
a_\rho^{-1} \simeq 3 a_\rho^{-1}$. 

\subsection{Free expansion and scaling} 

Now let us suppose that the trapping potential is switched-off at
$t=0$.  One can introduce a rescaled order parameter $\tilde\Psi$ 
as \cite{kagan}
\begin{equation}
\Psi(\rho,z,t)  = \frac{1}{b(t)} \tilde{\Psi} (\frac{\rho}{b(t)},z,t) 
\exp \left[ \frac{i m \rho^2 {\dot b}(t)}{2\hbar b(t) }
- \frac{i \mu \tau(t)}{\hbar} \right] 
\label{eq:Psitilde}
\end{equation}
where $\tau$ is defined as $\tau(t)=\int_0^t dt'/b^2(t')$. The scaling 
parameter $b(t)$ obeys the equation 
\begin{equation}
\ddot{b}(t)  =  \omega_\rho^2/b^3(t) \; , 
\label{eq:ddotb}
\end{equation}
with $b(0)=1$ and $\dot{b}(0)=0$, whose solution is 
\begin{equation}
b(t)=[1+\omega_\rho^2 t^2]^{1/2} \; .  
\label{eq:boft}
\end{equation}
With this choice, the GP equation (\ref{eq:TDGP}) for the rescaled 
order parameter $\tilde\Psi (\rho,z,t)$ becomes
\begin{eqnarray}
i \hbar \partial_t \tilde\Psi  &=&  \frac{1}{b^2(t)} 
\left( -\frac{\hbar^2\nabla_\rho^2}{2m}  +   V + 
\frac{gN}{L}|\tilde\Psi |^2 - \mu  \right) \tilde\Psi \nonumber \\
&-&  \frac{\hbar^2}{2m} \frac{\partial^2}{\partial z^2} \tilde\Psi \; . 
\label{eq:scaledGP}
\end{eqnarray}

A major property of this equation is that, if the rescaled 
order parameter at $t=0$ coincides with the $t$- and $z$-independent 
solution of the stationary GP equation (\ref{eq:Psi0}), then it remains 
$t$- and $z$-independent solution of the same equation forever. Let
us call $\tilde{\Psi}_0(\rho)$ this stationary solution. The time 
evolution of true order parameter $\Psi$ is entirely accounted for by 
the scaling parameter $b(t)$ given in (\ref{eq:boft}).  The stationary 
$\tilde\Psi_0(\rho)$ thus corresponds to an expanding order 
parameter with constant shape in the rescaled co-ordinate $\rho/b(t)$
and with a density that decreases in time as $n(\rho,t)= (N/L) 
\Psi_0^2(\rho/ b(t)) / b^2(t)$. This behavior is a particular case
of a more general scaling property of the GP equation, which is exact 
in a two-dimensional geometry and is valid for a generic 
time-dependent trapping 
frequency $\omega_\rho(t)$ \cite{kagan,castin,pitaevskii2}. 
Here we want to apply Eq.~(\ref{eq:scaledGP}) to describe also an 
expanding condensate whose initial configuration, at $t=0$, is a 
superposition of the ground state and some excited states.

\subsection{Rescaled Bogoliubov-like equations}

Let us take the rescaled order parameter in the form $\tilde\Psi
(\rho,z,t) = \tilde\Psi_0 (\rho) + \delta \tilde\Psi (\rho,z,t)$, 
where $\tilde\Psi_0$ is the (real) solution of the stationary GP 
equation and $\tilde{n}_0(\rho) = (N/L) \tilde{\Psi}_0^2(\rho)$ 
is the rescaled density. The linearized version of 
Eq.~(\ref{eq:scaledGP}) is
\begin{eqnarray}
i \hbar \partial_t \delta \tilde\Psi(\rho,z,t) & = & 
\left[ \frac{\tilde{H}_\rho}{b^2(t)}   
-\frac{\hbar^2\partial_z^2 }{2m} \right] \delta\tilde\Psi(\rho,z,t)
\nonumber \\ 
& + & \frac{g\tilde{n}_0(\rho)}{b^2(t)} \delta\tilde\Psi^*(\rho,z,t) \; ,
\label{eq:scaleddeltapsi}
\end{eqnarray}
with
\begin{equation}
\tilde{H}_\rho = -\frac{\hbar^2\nabla_\rho^2}{2m} + V(\rho) + 
2 g\tilde{n}_0(\rho)-\mu \; . 
\label{eq:Hrhotilde}
\end{equation}
Now let us write $\delta \tilde\Psi$ in the form 
\begin{equation}
\delta \tilde\Psi (\rho,z,t) = L^{-1/2} [ 
\tilde{u}(\rho,t) e^{ikz}+ \tilde{v}^*(\rho,t) e^{-ikz} ] \; .
\end{equation}
Inserting this expression into Eq.~(\ref{eq:scaleddeltapsi}) 
one gets the following equations for new amplitudes $\tilde{u}$ and
$\tilde{v}$ 
\begin{equation}
i\hbar\partial_t 
\left( \begin{array}{c} \tilde{u} \\ \tilde{v} \end{array} \right)
= \left[ \frac{1}{b^2(t)} {\cal L}_\rho + {\cal L}_z \right] 
\left( \begin{array}{c} \tilde{u} \\ \tilde{v} \end{array} \right)
\label{eq:scaledBG1}
\end{equation}
where
\begin{equation}
{\cal L}_\rho  = \left( \begin{array}{cc} \tilde{H}_\rho & 
g \tilde{n}_0 \\ -g \tilde{n}_0  & - \tilde{H}_\rho \end{array} \right)  
\; \; ; \; \;  
{\cal L}_z = \frac{\hbar^2 k^2}{2m} 
\left( \begin{array}{cc} 1 & 0 \\
0  & -1 \end{array} \right)  \; . 
\label{eq:scaledBG2}
\end{equation} 

These equations describe the behavior of small deviations from the
stationary rescaled order parameter $\tilde\Psi_0$. At $t<0$, when 
$b(t)=1$ and $\tilde\Psi=\Psi$, the $t$- and $\rho$-dependence 
of $\tilde{u}$ and $\tilde{v}$ can be factorized as in 
Eq.~(\ref{eq:linearized}) and the rescaled Bogoliubov-like 
equation (\ref{eq:scaledBG1}) reduces to the Bogoliubov equation 
(\ref{eq:bog}) for the excitations of the trapped condensate. One can 
select one of these excitations, with given $k$ and $n$, as initial 
values of $\tilde{u}$  and $\tilde{v}$  and then solve numerically 
Eq.~(\ref{eq:scaledBG1}) for the free expansion at $t>0$. Typical 
examples are given in Figs.~\ref{fig:uandv1} and \ref{fig:uandv2}, where 
we show the time evolution of $|\tilde{u}(\rho)|$ and $|\tilde{v}(\rho)|$ 
for six different excited states. 

\begin{figure}[htb]
\includegraphics[width=3.3in]{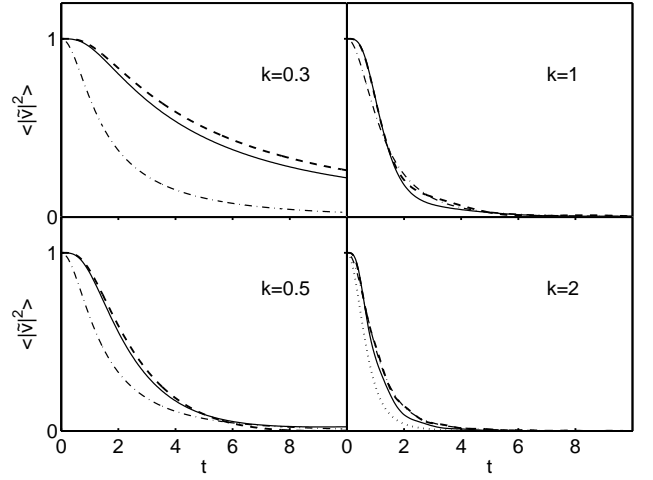}
\caption{ \label{fig:uandvmod2}
Normalized radial average of  $|\tilde{v}|^2$ as a function of $t$, in 
units of $\omega_\rho^{-1}$, for different values of $k$, in units of
$a_\rho^{-1}$. Solid lines
are the results obtained from Eq.~(\protect\ref{eq:scaledBG1}) for
the infinite cylinder. Dashed lines are the results obtained from 
Eq.~(\protect\ref{eq:kneq0}) or, equivalently, from 
Eq.~(\protect\ref{eq:schro}) and correspond to the behavior of 
$|\tilde{v}|^2$ in a uniform gas whose density is equal to the 
radial average of the density of the expanding infinite cylinder.  
Dot-dashed lines correspond to the adiabatic following approximation 
(\protect\ref{eq:adiabatic}). For larger values of $k$ all curves
converge to the analytic result (\protect\ref{eq:v2highk}) shown 
as a dotted line in the bottom-right quadrant. } 
\end{figure}

During the expansion $\tilde{u}$ and $\tilde{v}$ exhibit different 
behaviors: $\tilde{u}$ decreases in time at low $k$, while it remains 
almost constant at large $k$; vice-versa, $\tilde{v}$ always decreases, 
but it decreases faster at high $k$.  Let us define the normalized
radial average  $\langle |\tilde{v}|^2 \rangle = \int\! d\rho\ \rho 
|\tilde{v}(\rho,t)|^2 / \int d\rho \rho |v(\rho,0)|^2$. In 
Fig.~\ref{fig:uandvmod2} we plot $\langle |\tilde{v}|^2 \rangle$ as a 
function of $t$ for different $k$ (solid lines). At large $k$, all 
curve approach an universal behavior:  $\tilde{v}$ decreases in
the typical timescale of the expansion, $\omega_\rho^{-1}$, which 
is the time needed for the mean-field energy to vanish. Conversely, 
at low $k$ it decreases much more slowly, approaching a finite value 
for $t\to \infty$. This is also evident in Figs.~\ref{fig:uandv1} 
and \ref{fig:uandv2}, where the function $|\tilde{v}|$ for the
lowest value of $k$ remains well visible also at $t=5$ and $t=10
\omega_\rho^{-1}$, when the mean-field interaction is certainly 
negligible. In this regime, the functions $\tilde{u}$ and $\tilde{v}$ 
have completely lost their initial meaning of quasiparticle amplitudes 
of Bogoliubov modes; they are simply the two Fourier components of a 
density and velocity modulation in the expanding ``ideal" gas.

\begin{figure}[htb]
\includegraphics[width=3.3in]{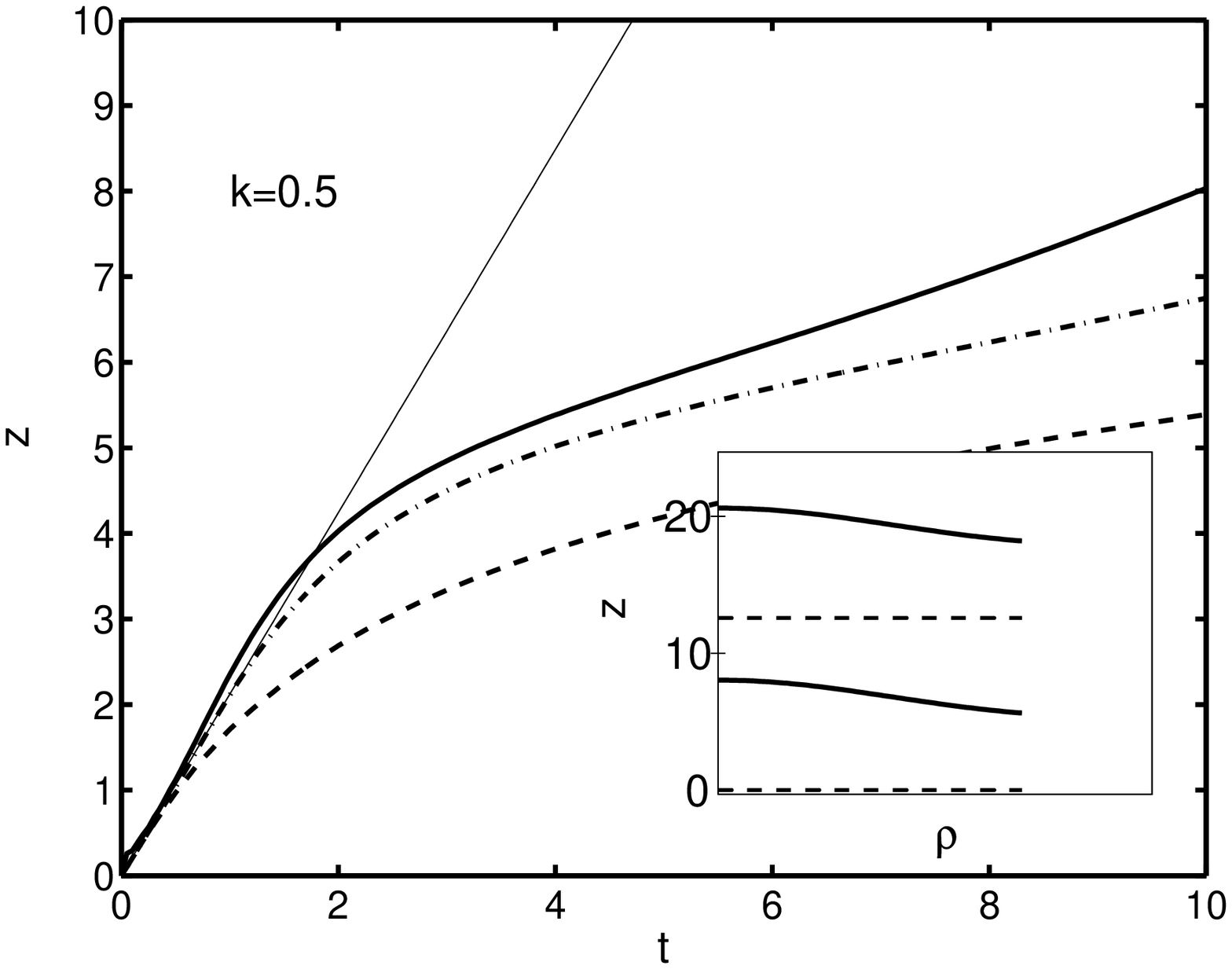}
\includegraphics[width=3.3in]{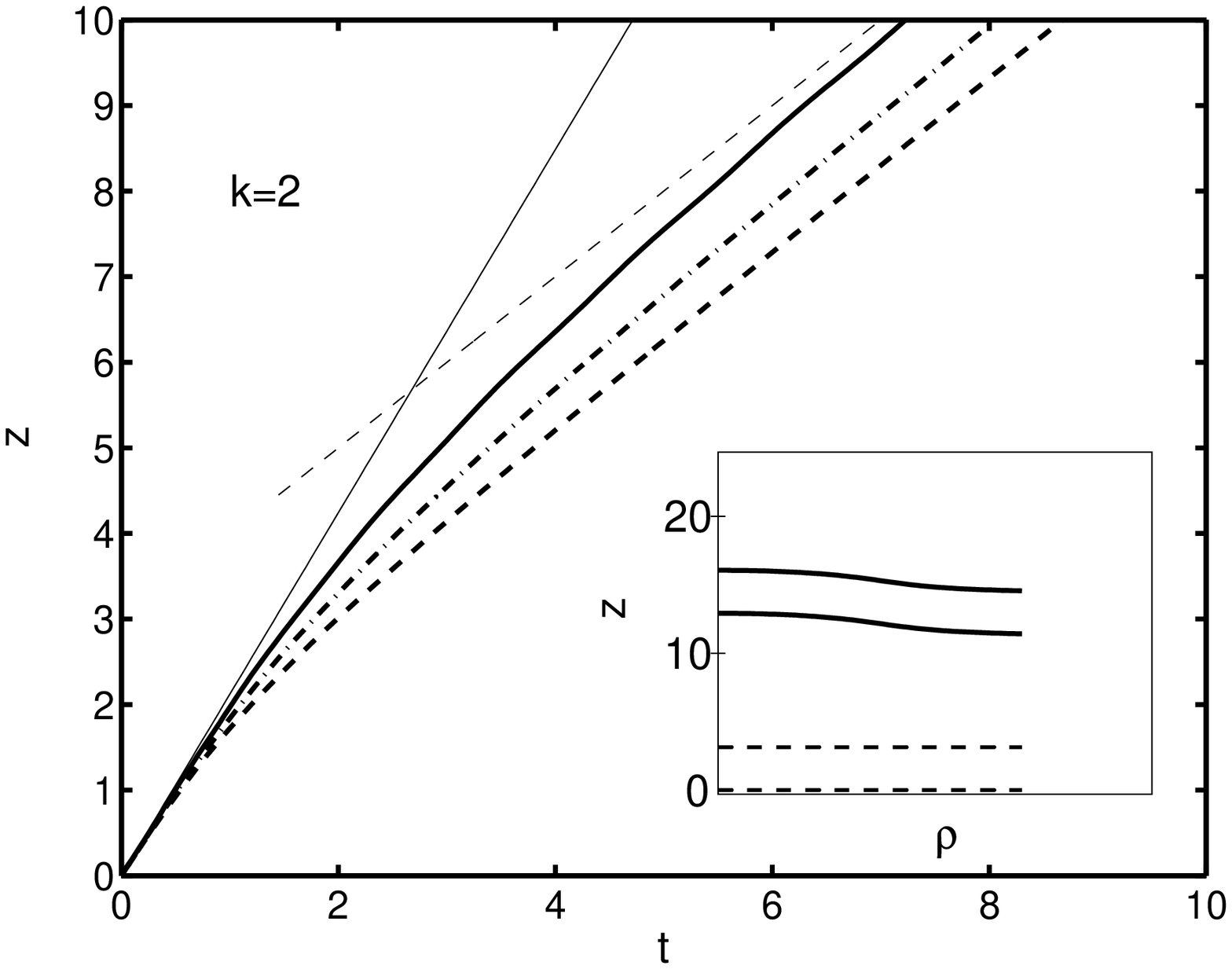}
\caption{ \label{fig:wavefront} 
Wavefront position {\it vs.} time for $k=0.5 a_\rho^{-1}$ and
$k=2 a_\rho^{-1}$, obtained from the maxima of $\delta\tilde{n}$ 
defined in Eq.~(\protect\ref{eq:deltatilden}) with $\tilde{u}$
and $\tilde{v}$ given by Eq.~(\protect\ref{eq:scaledBG1}). As in 
Fig.~\protect\ref{fig:GPfrontspeed}, the wavefront position is 
plotted for $\rho = 0$ (solid), $0.4$ (dot-dashed) and $0.8 R$ 
(dashed), where $R$ is the Thomas-Fermi radius of the condensate.
The slope of the thin solid and dashed lines is the Bogoliubov 
sound velocity, $c= [g n(0)/(2m)]^{1/2}$ and the phase velocity 
of free particles, $\hbar k /(2m)$, respectively. The inset shows 
the shape of two nearest fronts at $t=0$ (dashed lines) and
$t=10 \omega_\rho^{-1}$ (solid lines). Each front is plotted from
$\rho=0$ to $\rho=R=4.27 a_\rho$. The distance between the two 
fronts at $t=0$, in units of $a_\rho$, is $2\pi/k$.   }
\end{figure}

 From the knowledge of $\tilde{u}$ and $\tilde{v}$ one can also write 
the rescaled density variation $\delta \tilde{n}$, which is given by 
\begin{equation}
\delta \tilde{n} =  \tilde{\Psi}_0 |\tilde{u}+ \tilde{v}|
\cos[kz+\mbox{phase}(\tilde{u}+\tilde{v})] \; ,
\label{eq:deltatilden} 
\end{equation}
while $(\tilde{u} - \tilde{v})$ is related to the axial current 
density.  The initial phase can be chosen in such a way that the 
excitation at $t=0$ is just a sinusoidal wave of wavelength $2\pi/k$, 
which moves along $z$ with phase velocity $\omega_{nk}/k$. Then one 
can look at the time evolution of nodal lines or crests. Typical 
results are shown in Fig.~\ref{fig:wavefront}. The results at 
$k=0.5 a_\rho^{-1}$ can be compared with those obtained from the 
GP simulations in the elongated condensate for the same $k$, given
in Fig.~\ref{fig:GPfrontspeed}. The qualitative agreement between 
the two calculations is noticeable, especially if one keeps 
in mind that the results of Fig.~\ref{fig:wavefront} completely 
ignore the inhomogeneity and finite axial size of the elongated 
condensate of Fig.~\ref{fig:GPfrontspeed}.  The $k=2 a_\rho^{-1}$ 
case is also interesting, because it shows that the velocity 
of the fronts quickly 
approaches the phase velocity of free particles $\hbar k/(2m)$. 
Finally, from this results it is also clear that the observation of 
wavefronts, with their phase and amplitude, gives access to the 
quantity $|\tilde{u}+ \tilde{v}|$, while the total momentum 
transferred is a measure of $(|\tilde{u}|^2 - |\tilde{v}|^2)$ 
\cite{tozzo}. It is worth noticing that Eq.~(\ref{eq:scaledBG1}) 
conserves the total momentum, which is proportional to $\int\! 
d\rho \ \rho (|\tilde{u}|^2 - |\tilde{v}|^2)$. 

\subsection{Radial motion: scaling of nodal lines}

Now we consider in more details the time evolution of the radial 
shape of $\tilde{u}$ and $\tilde{v}$. In the two opposite limits
$k =0$ and $k\to \infty$ they obey an exact scaling law. Let us 
consider first the case $k=0$, i.e., purely radial excitations 
with an arbitrary number of radial nodes. The quantities $\tilde{u}$ 
and $\tilde{v}$ are $z$-independent and Eq.~(\ref{eq:scaledBG1})  
becomes
\begin{equation}
i\hbar\partial_t 
\left( \begin{array}{c} \tilde{u} \\ \tilde{v} \end{array} \right)
= \frac{1}{b^2(t)} {\cal L}_\rho 
\left( \begin{array}{c} \tilde{u} \\ \tilde{v} \end{array} \right) .
\label{eq:k=0}
\end{equation}
As initial condition at $t=0$, one can choose $\tilde{u} 
(\rho,0) = u_{n0}(\rho)$ and  $\tilde{v} (\rho,0) =  v_{n0}(\rho)$, 
where $u_{n0}$ and $v_{n0}$ are the eigenfunctions of the operator 
${\cal L}_\rho$ with eigenvalues $\hbar \omega_{n0}$, i.e., the 
$k=0$ solutions of the Bogoliubov eigenvalue problem (\ref{eq:bog}).  
Then an exact solution of Eq.~(\ref{eq:k=0}) is found in the 
form $\tilde{u} (\rho,t) = \alpha(t) u_{n0}(\rho)$ and  
$\tilde{v} (\rho,t) = \alpha(t) v_{n0}(\rho)$, with $\alpha(0)=1$. 
One finds 
\begin{equation}
i b^2(t) \partial_t \alpha(t) = \omega_{n0} \alpha(t) \; , 
\label{eq:alpha}
\end{equation}
whose solution is $\alpha(t) 
= \exp [- i \omega_{n0} \tau(t)]$ with $\tau(t) = \int_0^t dt'/[1+
\omega_\rho^2 t^{\prime 2}] = \omega_\rho^{-1} {\rm atan} (\omega_\rho t)$. 
The phase of $\tilde{u}$ and $\tilde{v}$ has a nontrivial $t$-dependence,
corresponding to a true oscillation of the type $\exp (i \omega_{n0} t)$ 
only for a short time interval, shorter than the typical expansion 
time $\omega_\rho^{-1}$, while for longer times the oscillations are 
frozen. However, the modulus $|\tilde{u}|$ and $|\tilde{v}|$ remains 
stationary at all times and the radial nodes of these functions remain 
fixed. This means that the density variation $\delta n$ has 
nodes that expand radially with the same scaling law of the whole 
condendate, governed by $b(t)$.   

The same scaling is found for $k\to \infty$ when, since $\tilde{v}$ 
is vanishingly small at all times,  Eq.~(\ref{eq:scaledBG1}) reduces 
to an equation for the amplitude $\tilde{u}$ only: $i \hbar b^2(t)
\partial_t \tilde{u} = [\tilde{H}_\rho + \hbar^2 k^2 /(2m) ] \tilde{u}$.  
If $\tilde{u}$ at $t=0$ is an eigenfunction $u_{nk}$ of $[\tilde{H}_\rho 
+ \hbar^2 k^2 /(2m) ]$, then it scales as $\tilde{u} (\rho,t) = 
\exp [- i \omega_{nk} \tau(t)] u_{nk}(\rho)$, as before.  

In the general case of a finite and nonzero value of $k$, the radial
and axial degrees of freedom in Eq.~(\ref{eq:scaledBG1}) are coupled
and the functions $|\tilde{u}(\rho)|$ and $|\tilde{v}(\rho)|$ are 
no more stationary. However, as one can see in Figs.~\ref{fig:uandv1}
and \ref{fig:uandv2}, their time evolution is very slow, at least for
the Bogoliubov modes with few radial nodes or none, and the positions 
of maxima and minima are almost constant. This is particularly 
evident for the minimum of $|\tilde{u}|$ in Fig.~\ref{fig:uandv2}, 
which remains always very close to the position of the radial node
at $t=0$ (solid circle on the axis).

\subsection{Axial motion: Thomas-Fermi approximation and local 
wavefront velocity}

\begin{figure}[htb]
\includegraphics[width=3.3in]{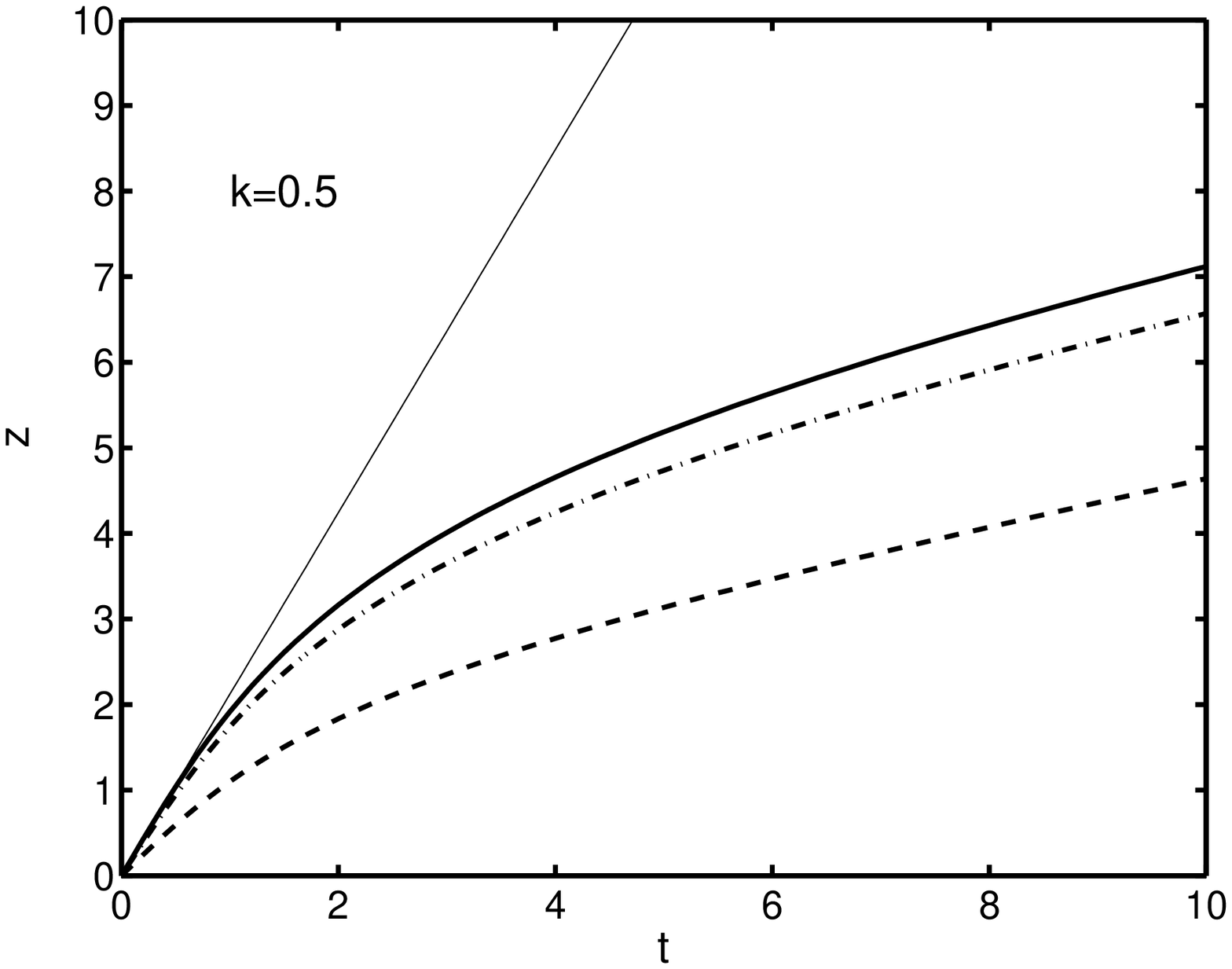}
\includegraphics[width=3.3in]{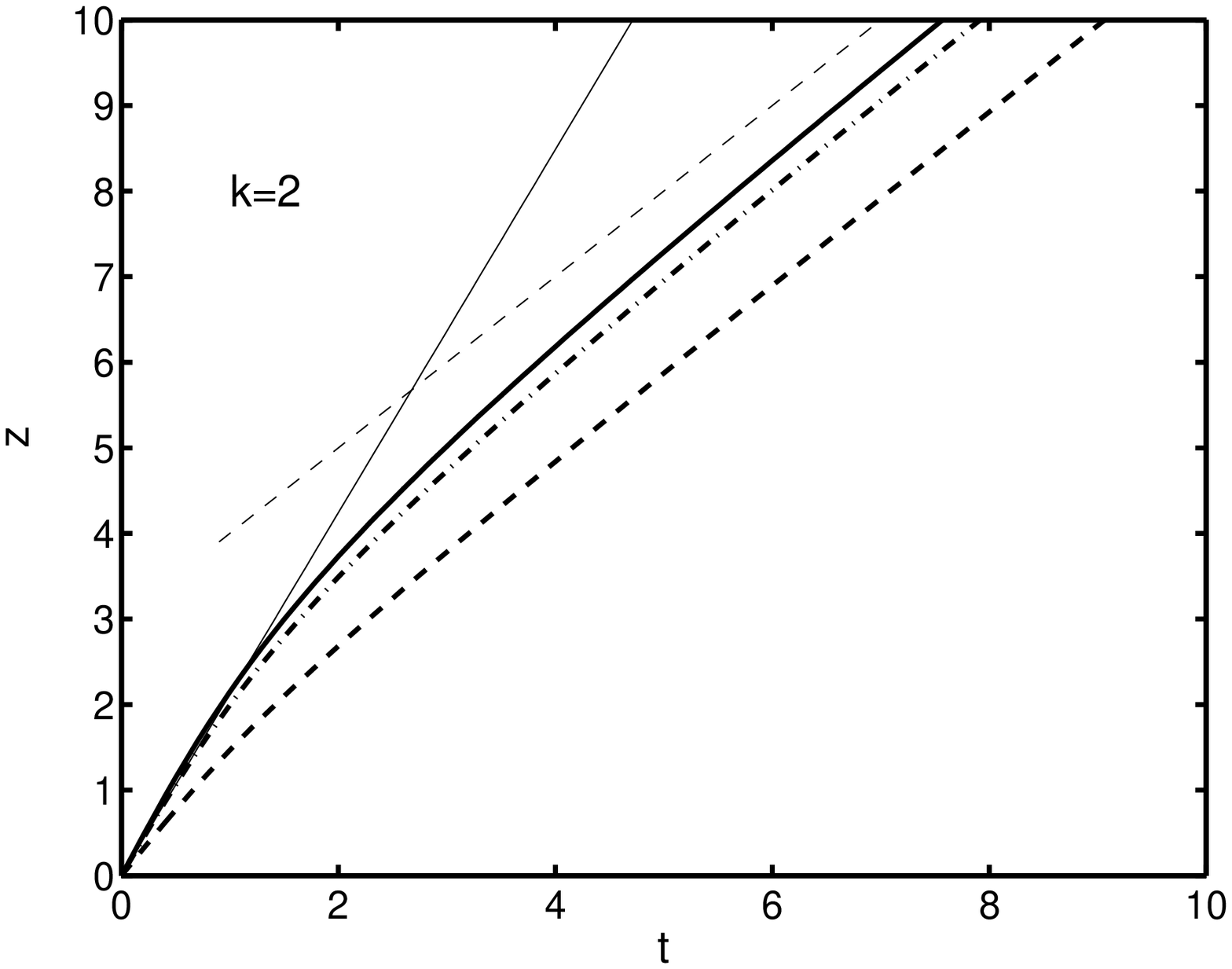}
\caption{ \label{fig:wavefrontTF}
Same as in Fig.~\protect\ref{fig:wavefront}, but using the local 
velocity (\protect\ref{eq:localspeed}) to get the wavefront position
at each $\rho$ and $t$.  }
\end{figure}

Here we concentrate on the axial motion of quasiparticles. We first 
observe that when $\eta \gg 1$ the ground state density of the 
infinite cylinder is well reproduced by the Thomas-Fermi 
approximation $g n_0(\rho)= \mu_{\rm TF} - V(\rho)$, which corresponds
to neglecting the  quantum pressure $-[\hbar^2/(2m)] \nabla_\rho^2$
in the stationary GP equation (\ref{eq:Psi0}). This also implies that 
$g \tilde{n}_0(\rho)= \mu_{\rm TF} - V(\rho)$ during the expansion. 
We also notice that the amplitudes $\tilde{u}$ and $\tilde{v}$ of axial 
phonons with a small number of radial nodes are smooth functions of 
$\rho$ at all times. This means that the term $[\hbar^2/(2m)] 
\nabla_\rho^2$ can safely be neglected also in Eq.~(\ref{eq:scaledBG1}), 
if we are not interested in the details of the radial motion. With 
these approximations, expression (\ref{eq:Hrho}) gives 
$\tilde{H}_\rho= g\tilde{n}_0$ and Eq.~(\ref{eq:scaledBG1}) 
becomes
\begin{equation}
i\hbar\partial_t 
\left( \begin{array}{c} \tilde{u} \\ \tilde{v} \end{array} \right)
 = H(t) \left( \begin{array}{c} \tilde{u} 
\\ \tilde{v} \end{array} \right) 
\label{eq:kneq0}
\end{equation}
where
\begin{equation}
H(t)  =  \frac{g \tilde{n}_0}{b^2(t)} \left( \begin{array}{cc} 1 & 1 \\ 
-1  & -1 \end{array} \right)   +  \frac{\hbar^2 k^2}{2m} 
\left( \begin{array}{cc} 1 & 0 \\ 0  & -1 \end{array} \right) \; .
\label{eq:Ht}
\end{equation}
The same equation can be written for the quantities $\tilde{u} +
\tilde{v}$ and $\tilde{u} - \tilde{v}$ \cite{maksimov}: 
\begin{eqnarray}
i \hbar \partial_t (\tilde{u}+\tilde{v}) & = &
 \frac{\hbar^2 k^2}{2m}  (\tilde{u}-\tilde{v}) \\
i \hbar \partial_t (\tilde{u}-\tilde{v}) & = &
\left( \frac{2 g \tilde{n}_0}{b^2(t)} + \frac{\hbar^2 k^2}{2m} \right)
 (\tilde{u}+\tilde{v}) \; . 
\end{eqnarray}
These equations can be easily decoupled to get the equation of motion
for $\tilde{u} + \tilde{v}$:
\begin{equation}
\partial_t^2 (\tilde{u} + \tilde{v}) = - \Omega^2(t) 
(\tilde{u} + \tilde{v})
\label{eq:densitywaves}
\end{equation}
with
\begin{equation}
\Omega^2(t) = \frac{k^2}{2m} \left( \frac{2g \tilde{n}_0}{b^2(t)} 
+ \frac{\hbar^2 k^2}{2m} \right) \; . 
\label{eq:Omega}
\end{equation}
The quantity $\Omega$ is the frequency of Bogoliubov excitations in
a uniform gas having time-dependent density $\tilde{n}_0/b^2(t)$.  
Recalling definition (\ref{eq:deltatilden}), one can identify the 
quantity
\begin{equation}
\tilde{c} = \frac{\Omega}{k} = \left[ \frac{g \tilde{n}_0}{m b^2(t)} + 
\left(\frac{\hbar k}{2m}\right)^2 \right]^{1/2} . 
\label{eq:localspeed}
\end{equation}
with the axial phase velocity of density modulations associated with
the expanding phonons. This velocity depends on $\rho$ through 
$\tilde{n}_0(\rho)$. One can insert the Thomas-Fermi density 
profile of the infinite cylinder and the expression of $b(t)$ to 
get analytic results for the wavefront position {\it vs.} time. 
Typical results of this local density approximation are shown in 
Fig.~\ref{fig:wavefrontTF} and can be compared with those obtained
from the numerical integration of the rescaled Bogoliubov equations,
in Fig.~\ref{fig:wavefront}. There is a good qualitative agreement both
at low and high $k$, even though, as expected, the local density 
approximation overestimates the $\rho$-dependence of $\tilde{c}$, 
especially for the motion near the surface of the condensate, where 
Thomas-Fermi approximation fails. 

\section{Adiabatic quasiparticle-to-particle evaporation}
\label{sec:twolevel}

In Eqs.~(\ref{eq:kneq0})-(\ref{eq:Ht}) the quantities $\tilde{n}_0$, 
$\tilde{u}$ and $\tilde{v}$ depend on $\rho$, but one 
can extract useful results also by averaging out this weak 
$\rho$-dependence. A simple approach consists of taking a constant 
rescaled density $\tilde{n}_0$ equal to the radial average $\langle 
\tilde{n}_0 (\rho) \rangle_\rho = \tilde{n}_0(0)/2$. Then, one can choose 
the initial $\tilde{u}$ and $\tilde{v}$ as the $\rho$-independent 
quasiparticle amplitudes of phonons in a uniform gas of that density 
and for a given $k$. Finally, one can solve Eq.~(\ref{eq:kneq0}) for 
the time evolution of the rescaled amplitudes. The resulting behavior 
of the function $|\tilde{v}(t)|^2$, normalized to its value at $t=0$, is 
shown in Fig.~\ref{fig:uandvmod2} for different $k$ (dashed lines). The 
remarkable agreement with the results for $\langle |\tilde{v}|^2 \rangle$ 
obtained from the rescaled Bogoliubov equations (\ref{eq:scaledBG1}) 
tells us that the evolution of a quasiparticle in the inhomogeneous 
cylindrical condensate is very similar to the one in a uniform gas 
having a density that decreases in time as $b^{-2}(t) = (1+\omega_\rho^2 
t^2)^{-1}$.  

An interesting reformulation of the same problem is found by considering 
Eq.~(\ref{eq:kneq0}) as the evolution of a two level system governed 
by a nonhermitian time-dependent Hamiltonian. Let us rewrite 
Eq.~(\ref{eq:kneq0}) in the form
\begin{equation}
i\hbar\partial_t | \delta \tilde{\Psi} \rangle 
 = H(t)| \delta \tilde{\Psi} \rangle 
\label{eq:kneq0-2}
\end{equation}
with
\begin{equation}
| \delta \tilde{\Psi} \rangle = 
\left( \begin{array}{c} \tilde{u} \\ \tilde{v} \end{array} \right) \; . 
\label{eq:deltatildePsi}
\end{equation}
Then let us recall definition (\ref{eq:Omega}) and introduce the 
quantity $\Theta(t)$ such that $\hbar \Omega \sinh \Theta = 
g \tilde{n}_0 / b^2(t)$ and $\hbar \Omega \cosh \Theta = 
g \tilde{n}_0 / b^2(t) + \hbar^2 k^2/(2m)$. Thus  
\begin{equation}
{\rm tanh}\Theta(t) = g \tilde{n}_0  \left( g \tilde{n}_0 
+ \frac{\hbar^2 k^2}{2m}  b^2(t) \right)^{-1}
\label{eq:Theta}
\end{equation}
and Hamiltonian (\ref{eq:Ht}) can be rewritten as
\begin{equation}
H(t) = \hbar \Omega(t)
\left(
\begin{array}{r r} \cosh \Theta(t) & \sinh \Theta(t) \\ 
-\sinh \Theta(t) & -\cosh \Theta(t) \end{array}
\right) .
\end{equation}
This Hamiltonian is nonhermitian. At any given time $t$, it admits two 
real eigenvalues $\pm \Omega$ and the corresponding biorthonormal set of
eigenvectors $\{ | \pm \rangle_r , | \pm \rangle_l \}$, where the
"right" and "left" eigenvectors are defined by 
\begin{eqnarray}
H(t)|\pm\rangle_r & = & \pm \hbar \Omega(t) |\pm\rangle_r   \\
H^\dagger (t)|\pm\rangle_l & = & \pm \hbar \Omega(t) |\pm\rangle_l  \; \; .
\end{eqnarray}
A simple calculation yields
\begin{eqnarray}
|+\rangle_r &=& \frac{1}{\sqrt{2}} \left(\begin{array}{c} \displaystyle
\sqrt{\cosh\Theta+1} \\
-\sqrt{\cosh\Theta-1} \end{array} \right) \label{eq:+r} \\
|-\rangle_r &=& \frac{1}{\sqrt{2}} \left(\begin{array}{c}\displaystyle
-\sqrt{\cosh\Theta-1} \\
\sqrt{\cosh\Theta+1} \end{array} \right) \label{eq:-r} \\
_l\langle+| &=& \frac{1}{\sqrt{2}}\left(\begin{array}{c c}\displaystyle
\sqrt{\cosh\Theta+1},&
\sqrt{\cosh\Theta-1} \end{array}\right) \label{eq:+l} \\
_l\langle-| &=& \frac{1}{\sqrt{2}} \left(\begin{array}{c c}\displaystyle
\sqrt{\cosh\Theta-1},&
\sqrt{\cosh\Theta+1}\end{array}\right) \; . \label{eq:-l}
\end{eqnarray}
These vectors obey the orthogonality relations 
\begin{eqnarray}
_l\langle+|+\rangle_r &=& _l\langle -|-\rangle_r = 1 
\label{eq:biortho}\\
_l\langle+|-\rangle_r &=& _l\langle-|+\rangle_r = 0 \; .
\end{eqnarray}
They also satisfy the relations
\begin{eqnarray}
_l\langle\pm|\partial_t|\pm\rangle_r &=& 0 \\
_l\langle\mp|\partial_t|\pm\rangle_r &=& -(1/2)\partial_t\Theta(t)
\end{eqnarray}
with
\begin{equation}
\partial_t\Theta(t)= -\frac{\partial_t b^2(t)}{b^2(t)}
\frac{g \tilde{n}_0}{2g \tilde{n}_0+ \hbar^2 k^2 b^2(t)/(2m)} \; .
\label{eq:deltatheta}
\end{equation}

Notice that the $\tilde{u}$ and $\tilde{v}$ components of the
eigenvectors (\ref{eq:+r})-(\ref{eq:-l}) are such that $|\tilde{u}|^2 
- |\tilde{v}|^2 = \pm 1$ for $|\pm \rangle$ states, with eigenfrequency 
$\pm \Omega$. At $t=0$ the state $|+ \rangle_r$, corresponds to a 
quasiparticle of wavevector $k$ and frequency $\Omega(0) = \{ k^2/(2m) 
[2g \tilde{n}_0 +\hbar^2 k^2/(2m)]\}^{1/2}$, whose components, 
$\tilde{u}=u_k$ and $\tilde{v}=v_k$, are the usual amplitudes of a 
Bogoliubov mode in a uniform gas. At $t \to \infty$ the same state 
corresponds to a free atom with the same $k$ and frequency 
$\Omega(\infty)=\hbar k^2/(2m)$. The time evolution of the state 
$|+ \rangle_r$ is simply obtained by using the expression 
(\ref{eq:+r}) and the definitions of $\Theta(t)$ and $\Omega(t)$. 
For example, one gets
\begin{equation}
\tilde{v}^2 (t) = \frac{g\tilde{n}_0/b^2(t) + \hbar^2 k^2/(2m)}{ 2
\hbar \Omega(t)} - \frac{1}{2} \; ,
\label{eq:adiabatic}  
\end{equation}
which has the two correct limits $\tilde{v}^2(0)= v_k^2 =
[gn_0 + \hbar^2 k^2/(2m)]/(2\hbar \Omega(0)) - 1/2$ \cite{ps} and 
$\tilde{v}^2(\infty)=0$.  In the large $k$ limit, when  
$\hbar^2 k^2/(2m) \gg gn_0$, Eq.~(\ref{eq:adiabatic}) gives 
\begin{equation}
\tilde{v}(t) = \frac{m g\tilde{n}_0}{\hbar^2 k^2 b^2(t)} \; . 
\label{eq:v2highk}  
\end{equation}

Now, let us use the basis of instantaneous eigenvectors 
$\{ |\pm \rangle_r \}$ to project the state $|\delta \Psi\rangle$ as 
\begin{equation}
|\delta \Psi\rangle = c_+ |+\rangle_r + c_- |-\rangle_r
\end{equation}
and assume that the initial state is just a quasiparticle of 
momentum $\hbar k$, that is $c_+=1$ and $c_-=0$ at $t=0$. The 
evolution of  $|\delta \Psi\rangle$ is governed by the Schr\"odinger 
equation (\ref{eq:kneq0-2}). By using the new basis together with  
the properties (\ref{eq:biortho})-(\ref{eq:deltatheta}), the
same equation can be rewritten in the form 
\begin{equation}
\begin{array}{c} i\partial_t\\ \end{array}
\left(\begin{array}{c}c_+(t)\\c_-(t)\end{array}\right)
= \left(\begin{array}{c c}
\Omega(t) & \frac{i}{2}\partial_t\Theta(t) \\
\frac{i}{2}\partial_t\Theta(t) & - \Omega(t) \end{array}\right)
\left(\begin{array}{c}c_+(t)\\c_-(t)\end{array}\right) \; . 
\label{eq:schro}
\end{equation}
The numerical integration of this equation gives, of course, the 
same results of Eq.~(\ref{eq:kneq0-2}) (dashed lines for 
$|\tilde{v}|^2$ in Fig.~\ref{fig:uandvmod2}, for instance), but the 
use of a different basis makes the evaporation mechanism more 
transparent. In fact, it allows one to point out the interesting 
situation that occurs when $|\delta \Psi\rangle$ remains 
close to $|+\rangle_r$ at all times. This {\it adiabatic following} 
happens when the diabatic coupling between the two eigenstates is 
small, i.e., when the off-diagonal term $|\partial_t \Theta|$ is 
much smaller than the frequency splitting $2\Omega$ \cite{messiah}. 
Now, the function $|\partial_t\Theta(t)|$ vanishes at $t=0$ and 
$t \to \infty$ and is always smaller than $\partial_t b^2/(2b^2)$ 
which has maximum value $\omega_\rho/2$ for $t=\omega_\rho^{-1}$. 
Conversely, the function $\Omega(t)$ decreases monotonically from 
$\Omega(0)$ to $\hbar k^2/(2m)$. Thus the evolution is certainly 
adiabatic if $\hbar k^2/(2m)\gg \omega_\rho/2$, that means
$k \gg a_\rho^{-1}$. This condition is somewhat too restrictive.
In fact, the two functions $|\partial_t \Theta|$ and $2\Omega$ might 
be comparable only around $t=\omega_\rho^{-1}$ and the adiabatic 
evolution is ensured if $\Omega \gg \omega_\rho$ at that time. Now,
for $k$ of the order of $a_\rho^{-1}$ and $\eta = g \tilde{n}_0/ 
(\hbar \omega_\rho) \gg 1$, the term  $g \tilde{n}_0$ in $\Omega$ 
is larger than $\hbar^2 k^2/(2m)$ at $t=0$, but it remains larger 
also at $t=\omega_\rho^{-1}$, when $b^2=2$. The condition for 
adiabaticity thus becomes $k \gg \eta^{-1/2} a_\rho^{-1}$. In our
case $\eta=9.1$ and the condition is $k \gg 0.3 a_\rho^{-1}$.

\begin{figure}[htb]
\includegraphics[width=3.3in]{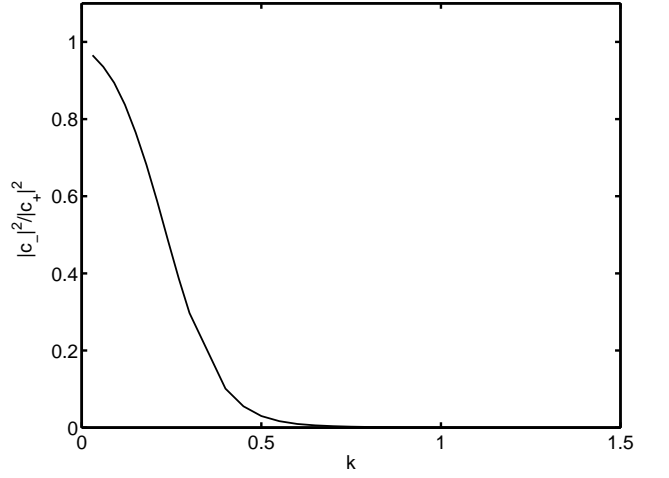}
\caption{ \label{fig:c-}
Ratio $|c_-|^2/|c_+|^2$ at $t\to \infty$ obtained by solving 
Eq.~(\protect\ref{eq:schro}) for different values of $k$, in units of
$a_\rho^{-1}$, with the condition $c_+=1$ and $c_-=0$ at $t=0$.  }
\end{figure}

One can check whether the quasiparticle-to-particle conversion 
is adiabatic or diabatic by directly comparing the evolution of the 
state $|\delta \Psi\rangle$ with that of the instantaneous eigenstate
$|+\rangle_r$. The difference is a measure of non-adiabaticity. 
For the quantity $\tilde{v}^2$ in Fig.~\ref{fig:uandvmod2} the 
comparison has to be done between dashed lines, that come from 
the evolution of $|\delta \Psi\rangle$ through Eq.~(\ref{eq:schro}),
and dot-dashed lines, that come from the evolution of $|+\rangle_r$ 
through Eq.~(\ref{eq:adiabatic}).  As one can see, for $k \gg 0.3 
a_\rho^{-1}$ the adiabatic following approximation is indeed an 
accurate description of the evaporation process. The dashed and 
dot-dashed curves are already indistinguishable for $k=2a_\rho^{-1}$ 
and, for larger $k$, all curves collapse on the dotted curve, which 
represents the asymptotic law (\ref{eq:v2highk}). For low $k$, 
conversely, the diabatic coupling between the $|+\rangle_r$ 
and $|-\rangle_r$ eigenstates, originating from the time-dependence
of the scaling parameter $b(t)$, is no more negligible and the
state $|\delta \Psi\rangle$ significantly differs from $|+\rangle_r$
both at short and long times. A nice view of this coupling can be seen 
in Fig.~\ref{fig:c-}, where we plot the value of $|c_-|^2/|c_+|^2$ 
at $t \to \infty$ as a function of $k$. In the adiabatic case 
this ratio vanishes.  At low $k$, conversely, it tends to $1$. 
From the definitions (\ref{eq:+r})-(\ref{eq:-r}), one can easily see 
that the quantity $|c_-|^2/|c_+|^2$ at $t=\infty$ is equal to 
$|\tilde{v}|^2/|\tilde{u}|^2$ and corresponds to the ratio between 
the number of atoms moving with opposite momenta $\pm \hbar k$ at 
$t \to \infty$, as a result of the evaporation of an initial 
quasiparticle with momentum $+\hbar k$. It should not be confused,
vice-versa, with the ratio $|v_k|^2/|u_k|^2$ at $t=0$, which 
is different and is related to $\pm k$ components the momentum 
distribution of the initial in-trap quasiparticle \cite{vogels,brunello}.

\section{GP simulations revisited}
\label{section3}

We are now ready to use the results of sections III and IV to 
interpret some relevant features of GP simulations in elongated 
condensates. 

\bigskip

{\it Axial motion of wavefronts}

Differently from the infinite cylinder, the elongated condensate
expands also in the $z$-direction. Let us consider 
Eq.~(\ref{eq:localspeed}) and apply a local density approximation 
in this form:
\begin{equation}
\tilde{c} (\rho,z,t)  = \frac{{\dot b}_z(t)}{b_z(t)} z +
\left[ \frac{ g \tilde{n}_0(\rho,z)}{m b_\rho^2(t)b_z(t)} 
+ \left(\frac{\hbar k}{2m b_z(t)}\right)^2 
\right]^{1/2} \, .  
\label{eq:simplemodel}
\end{equation}
The slow expansion along $z$ is here included through the scaling 
parameter $b_z$. The new scaling laws are \cite{kagan,castin,minniti}
\begin{equation}
\ddot{b}_\rho (t) = \frac{\omega_\rho^2}{b_\rho^3(t) b_z(t)}
\, \, \, ; \, \, \, 
\ddot{b}_z (t) = \frac{\lambda^2 \omega_\rho^2}{b_\rho^2(t) b_z^2(t)} . 
\label{eq:scaling2d}
\end{equation}
Thus the first term in $\tilde{c}$ is the local drift velocity
due to the axial motion of the ``background" in which the excitations 
move. The second term is now $\rho$- and $z$-dependent through 
the rescaled density $\tilde{n}_0(\rho,z)$. This simple model 
allows us to plot the wavefront position and speed for the 
elongated condensate, by using the unperturbed GP density profile 
at $t=0$ as input and solving the two coupled equations for 
$b_z$ and $b_\rho$. The position of the wavefronts, corresponding 
to the simulation in Fig.~\ref{fig:GPdeltan1}, are shown as dashed 
lines in the same figure. In practice, we take
the crests of $\delta n$ at $t=0$ from the GP simulation and let 
them move with velocity $\tilde{c}$ during the expansion for $t>0$. 
The dashed lines are the same crests at $t=25 \omega_\rho^{-1}$. 
They nicely agree with the results of GP simulations (solid lines),
except near the surface where the Thomas-Fermi approximation is
inadequate. 

\begin{figure}[htb]
\includegraphics[width=3.3in]{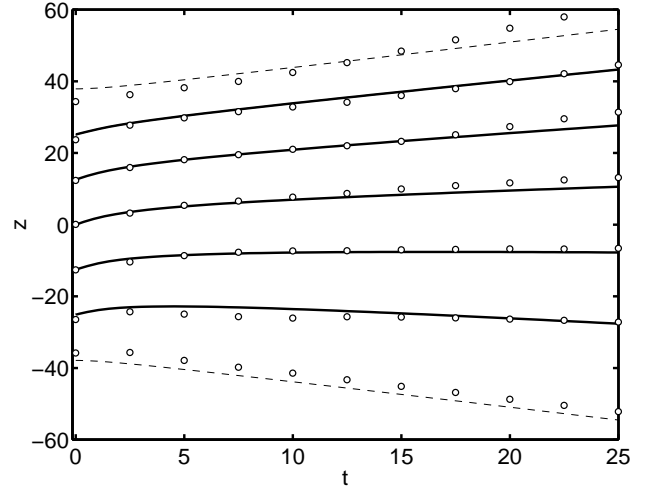}
\caption{ \label{fig:patterns}
Wavefront position {\it vs.} time for the elongated
condensate of Figs.~1 and 3. Points correspond to the crests of
the linear density obtained from the GP simulation. Solid lines 
are the results of the local density model, with local phase 
velocity given by (\protect\ref{eq:simplemodel}). Thin dashed 
lines represent the axial size of the expanding condensate. 
Length and time are in units of $a_\rho$ and $\omega_\rho^{-1}$,
respectively. 
}
\end{figure}

One can also estimate the position in time of the crests of the 
linear density (upper part of Fig~\ref{fig:GPdeltan1}). A simple 
way consists in propagating each front with a radially averaged 
velocity, which is obtained by replacing $\tilde{n}_0(\rho,z)$ 
in Eq.~(\ref{eq:simplemodel}) with the average $\langle 
\tilde{n}_0(\rho,z) \rangle_\rho$. The
results are shown as solid lines in Fig.~\ref{fig:patterns}, where
they are compared with the ones of the GP simulation (empty circles).
The thin dashed lines represents the Thomas-Fermi axial size of the 
expanding condensate.  The condensate has maximum density at $z=0$.
All crests move initially with almost the same positive velocity 
along $z$. Within a short time interval, when the mean-field 
is still active, those crests that move downhill (positive $z$)
tend to be accelerated by the expanding background, with
respect to their motion in an infinite cylinder. Those 
that move uphill (negative $z$) seem to be decelerated, and they 
can also invert their motion. 
The agreement with the full GP simulation is remarkable. This 
means that Eq.~(\ref{eq:simplemodel}) accounts for most of the 
physics involved in the axial motion of density modulations in 
the expansion of low $k$ phonons. 

\bigskip

{\it Threshold for the appearance of a released-phonon cloud}

Looking at Figs.~\ref{fig:wavefront} and \ref{fig:patterns} one
understands why at low $k$ the excited atoms do not form a separate 
cloud: they have not enough speed to reach the expanding boundary 
along $z$. In classical fluids this situation corresponds to the case 
of sound waves propagating on a background hydrodynamic flow. A 
boundary that moves faster than the speed of sound acts as  
an event horizon in a sonic black hole \cite{unruh}. 
A possible realization of a stable black hole for long wavelength 
phonons in ring-shaped Bose-Einstein condensates has been recently 
proposed in Ref.~\cite{garay}. Our configuration is nonstationary
and the excitations are sound-like waves only at the very beginning
of the expansion, in a time interval of the order of $\omega_\rho^{-1}$, 
when they are still well inside the condensate. Then, the wavefronts 
reach an almost constant phase velocity, of the order of $\hbar k/(2m)$ 
and the whole wave packet moves at group velocity $\sim \hbar k/m$. 
Thus, if the latter is smaller than the velocity of the boundary 
(upper dashed line in  Fig.~\ref{fig:patterns}), most of the atoms 
remain within the expanding condensate at all times. In the limit 
$\lambda \ll 1$, Eqs.~(\ref{eq:scaling2d}) yield the expression 
$(\pi/2) \omega_\rho \lambda^2  Z_{\rm TF}$ for the asymptotic 
velocity of the boundary, where $Z_{\rm TF}$ is its initial position 
in Thomas-Fermi approximation \cite{castin}.  Thus the condition for the 
atoms to exit is $\hbar k/m > (\pi/2) \omega_\rho \lambda^2 Z_{\rm TF}$. 
Since $(1/2)m \lambda^2 \omega_\rho^2 Z_{\rm TF}^2 = \mu_{\rm TF}$ 
and $\eta= \mu_{\rm TF}/(\hbar \omega_\rho)$, the same condition 
can be rewritten as $ ka_\rho > \pi \lambda (\eta/2)^{1/2}$. 
With the parameters of the condensate in section II, this threshold 
is around $k \sim 0.76 a_\rho^{-1}$, in agreement with the transition 
from the low $k$ scenario of Fig.~\ref{fig:GPdensity1} to the high 
$k$ scenario of Fig.~\ref{fig:GPdensity2}. The transition is not 
sharp, however, since the excited atoms can reach the boundary at 
different times and with different velocities so that, for a given 
expansion time, one can observe a released-phonon cloud only 
partially separated from the condensate.

\bigskip

{\it Adiabatic and diabatic evaporation}

For the elongated condensate of the simulations in section II one has 
$\eta^{-1/2} < \pi \lambda (\eta/2)^{1/2} < \eta^{1/2}$. The first value 
is the threshold for the adiabatic quasiparticle-to-particle evaporation 
discussed in section IV (see Fig.~\ref{fig:c-}). The second is the 
threshold for the appearance of a separate released-phonon cloud. The 
third is the inverse of the healing length. This implies that when the 
initial excitations are single-particle modes, i.e., with 
$ka_\rho > \eta^{1/2}$, the excited atoms always move out of the 
condensate. Instead, when the initial excitations are phonon-like modes, 
i.e., with $ka_\rho < \eta^{1/2}$, they can either move out or remain 
inside. However, if they move out as a separate cloud, the evaporation 
is certainly adiabatic. This means that a single quasiparticle of positive
momentum $\hbar k$ gives rise to a single atom moving with the 
same positive momentum, even though the initial quasiparticle state 
is made of correlated $+k$ and $-k$ components in the momentum 
distribution of the atoms. It is worth noticing that, with  
appropriate choices of $\lambda$ and $\eta$, such that $\pi \lambda 
(\eta/2)^{1/2} <  \eta^{-1/2}$, one can have diabatic evaporation
together with the appearance of released-phonon clouds. In this case,
one expects to have atoms moving with both positive and negative 
momentum $\pm \hbar k$ out of the condensate.   

\begin{figure}[htb]
\includegraphics[width=3.3in]{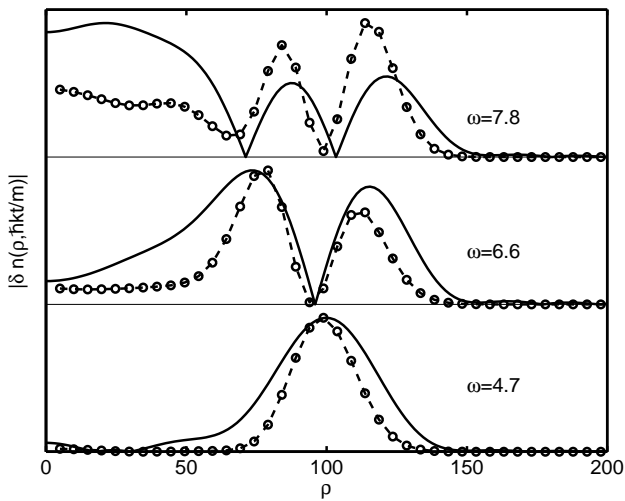}
\caption{ \label{fig:nodes}
Radial density profiles, in arbitrary units,  of the released-phonon 
cloud at $z=\hbar k t/m$, with $t=25 \omega_\rho^{-1}$, 
$k=2.5 a_\rho^{-1}$ and three different values of $\omega$. Points 
with dashed lines are the results of GP simulations with $V_B = 0.3 
\hbar \omega_\rho$ and $t_B = 4 \omega_\rho^{-1}$. Solid lines are 
the functions $|u_{nk}+v_{nk}|$ for an infinite cylinder with 
$\eta=9.1$, the same $k$ and $n=0$ (bottom), $1$ (mid) and $2$ (top). 
}
\end{figure}

\bigskip

{\it Radial scaling of nodal lines}

In section II, we have seen that the released-phonon cloud 
exhibits a ``shell" structure (see Figs.~\ref{fig:GPdensity2} and 
\ref{fig:GPdeltan2}). It can be reasonably assumed that the scaling 
of the radial motion, which is almost exact for the infinite 
cylinder, is still valid also for the elongated condensate. This
means that an excited state having a radial node at a point 
$\rho_0(z)$ at $t=0$ should produce a cloud of excited atoms 
having a minumum of density close to $\rho_0/b(t)$, but translated 
along $z$ as a consequence of the axial motion. If this is true,
the shell structure in  Figs.~\ref{fig:GPdensity2} and 
\ref{fig:GPdeltan2} is an indication that, even though we used a 
Bragg frequency in resonance with the lowest phonon branch with 
no nodes, some quasiparticles in the next branches, with one and 
two radial nodes, were also excited. In order to be more selective,
one has to use Bragg pulses with longer duration $t_B$ 
\cite{steinhauer2,tozzo}.  We thus repeat the same type of 
simulations of Figs.~\ref{fig:GPdensity2} and \ref{fig:GPdeltan2}, 
for $k=2.5 a_\rho^{-1}$, but with $t_B=4\omega_\rho^{-1}$. The
Bragg frequency is varied in such a way to excite phonons with $0$, 
$1$ and $2$ radial nodes. We find a significant change in the 
shell structure of the released-phonon cloud. In Fig.~\ref{fig:nodes}
we plot a cut of its density distribution at $z=\hbar k t/m$, 
with $t=25 \omega_\rho^{-1}$, which corresponds roughly to the 
position where the moving cloud has its maximum radial extension. 
The three density profiles obtained from GP simulations (points 
with dashed lines) are compared with the shape of the functions
$|u_{nk}+v_{nk}|$ (solid lines), calculated at $\rho/b(t)$, for 
the Bogoliubov modes of an infinite cylinder, with the same $k$ 
and with $n=0,1$ and $2$. The good agreement confirms that the 
nodal lines scale almost exactly and that the observation of radial 
structures in the released-phonon cloud can be an efficient tool for 
the identification of the initial in-trap quasiparticles.

\section{Conclusions}

We have shown that the evaporation of phonons in a freely expanding 
condensate is an interesting quantum process with a rich variety 
of scenarios. Our analysis is based on the GP theory at zero 
temperature and in the linear response regime, where the concept 
of quasiparticle is well defined. We have thus neglected possible 
effects of thermal and quantum fluctuations, as well as those of 
nonlinear dynamics. This is justified if one assumes that the 
population of the initially excited phonon-like mode is large enough 
to neglect fluctuations and small enough to remain in the linear regime. 
For condensates with $10^5$ to $10^6$ atoms or more, this 
condition is ensured by choosing the Bragg intensity and duration
such to excite about $1$ to $10$\% of the atoms as in typical experiment.  
In this paper, we have provided explicit calculations for an elongated
condensate similar to that of current experiments, but the results
obtained for the infinite cylinder are rather general. The extension 
of this analysis to different sets of parameters can lead to different 
regimes that are also worth exploring. 

\acknowledgments

We are indebted to N.~Davidson, N. Katz, R.~Ozeri, L.~Pitaevskii, 
P.O.~Fedichev, P.~Massignan, M.~Modugno and L.~Ricci for useful discussion. 
This work is supported by the Ministero dell'Istruzione, dell'Universit\`a 
e della Ricerca. One of the authors (FD) thanks the Dipartimento di 
Fisica di Trento for the hospitality.

\end{document}